\documentclass[useAMS,usenatbib,usegraphicx]{mn2e}

\title[Generation of density inhomogeneities by MHD waves in 2D]
{Generation of density inhomogeneities by magnetohydrodynamic waves in
two dimensions}

\author[S. Van Loo et al.]
{S. Van Loo,$^{1}$\thanks{E-mail: svenvl@ast.leeds.ac.uk} 
S. A. E. G.  Falle$^2$ and T. W. Hartquist$^1$\\
$^1$School of  Physics and Astronomy,  University of Leeds,  Leeds LS2
9JT\\  $^2$Department  of Applied  Mathematics,  University of  Leeds,
Leeds LS2 9JT}

\begin{document}

\date{Accepted - . Received - ; in original form -}

\pagerange{\pageref{firstpage}--\pageref{lastpage}} \pubyear{2006}

\maketitle

\label{firstpage}

\begin{abstract}
Using  two   dimensional  simulations,  we  study   the  formation  of
structures with  a high-density  contrast by magnetohydrodynamic waves 
in regions in which the ratio of thermal to magnetic pressure is small.  
The initial state is a uniform background perturbed by fast-mode  wave.  
Our most significant result is that dense  structures persist for far 
longer in a two-dimensional simulation than in the one-dimensional 
case. Once formed, these structures persist as long as the fast-mode 
amplitude remains high.
\end{abstract}


\section{Introduction}

One of the outstanding problems in star formation is the determination 
of the mechanisms that are responsible for the observed  clumpiness in 
star forming regions.  Since the ratio,  $\beta$,  of the thermal to
magnetic pressure is observed to be small \citep{CR99}, it is natural
to suppose that magnetohydrodynamic (MHD) waves play a crucial role. 
For this reason, there have been a number of three-dimensional (3D)
simulations of the effect of MHD waves on an isothermal plasma in 
which $\beta$ is small (e.g. \citealt{BM02};  \citealt{PN02}; 
\citealt{Getal03}; \citealt{Letal04}; \citealt{NL05};  \citealt{V05}). 
These do, indeed, show that density inhomogeneities can be formed in 
this way with statistical properties that are consistent with the 
observations. However, these calculations generally contain so many 
ingredients that they obscure the fundamental mechanisms.

In two previous papers (\citealt{FH02}, hereafter FH02 and
\citealt{LFH05},  hereafter  LFH05), we carried out 1D calculations of 
MHD waves in a low-$\beta$ plasma that were designed to elucidate the 
mechanism by which clumps are formed. In FH02, we showed that clumps 
with a high-density contrast could be generated by the excitation of 
slow-mode waves due to the non-linear steepening of a fast-mode wave.
In LFH05, we explored  how  these effects were modified by the 
dissipation due to ion-neutral friction. The present paper is an  
extension of the work in FH02 to two dimensions. We find that the 
mechanism described in the previous paper works even better in two  
dimensions in the sense that the density contrasts are both larger and 
persist for longer.

The  governing  equations  and  the  initial condition  are  given  in
Sect.~\ref{sect:numerical calculations}.  We  then discuss the results
for a specific  numerical model (Sect.~\ref{sect:results}) and examine
how  these depend on  the model  parameters (Sect.~\ref{sect:parameter
study}).  Finally,  we conclude the  paper by relating our  results to
observations     of    clumps    and     dense    cores     in    GMCs
(Sect.~\ref{sect:discussions and conclusions}).

\section{Numerical calculations}

\label{sect:numerical calculations}

\subsection{Governing equations}

While FH02  considered only variations  in the $x$-direction,  we also
allow changes in  the $y$-direction.  Since we are  only interested in
magnetosonic waves, there  is no need to introduce  a $z$-component of
either the  magnetic field or the  velocity. The MHD  equations for an
isothermal gas can then be written in the form

\[
\frac{\partial{{\bf       U}}}{\partial{t}}+       \frac{\partial{{\bf
F}}}{\partial{x}}+ \frac{\partial{{\bf G}}}{\partial{y}} = 0,
\]

\noindent
where 

\[
{\bf U} = [\rho, \rho v_x, \rho v_y, B_x, B_y]^t
\]

\noindent
is a vector of conserved variables, and 

\[
\begin{array}{lcl}
{\bf F} &  = & [\rho v_x, \rho  v_x^2 + p_g + B^2/2 -  B_x^2, \rho v_x
v_y - B_x B_y, 0, v_x B_y - v_y B_x]^t,\\
{\bf G} &  = & [\rho v_y, \rho v_y v_x - B_y B_x, \rho  v_y^2 + p_g +
B^2/2 -  B_y^2, v_y B_x - v_x B_y, 0]^t,\\
\end{array}
\]

\noindent
are the  corresponding fluxes.  Here  we assume that $p_g  =\rho a^2$,
where $a$ is a constant sound speed. These equations were solved using
the second order upwind scheme described in \citet*{FKJ98}, except that
we  use  the divergence cleaning  technique  described  by \citet{D02}.

\subsection{Initial state}

As  in  FH02,   we  are  interested  in  the   generation  of  density
inhomogeneities  by  fast-mode   waves  propagating  in  the  positive
$x$-direction.  Such a wave propagates with speed

\[
v_x + c_f,
\]

\noindent
where the fast magnetosonic speed, $c_f$, is given by 

\[
c_f^2  =  \frac{1}{2}  \left[B^2/\rho+a^2 +  \sqrt{(B^2/\rho+a^2)^2  -
4a^2B_x^2/\rho}\right].
\]

The form of such a wave is most conveniently written in terms of the
primitive variables

\[
{\bf P} = [\rho, v_x, v_y, B_x, B_y]^t.
\]

\noindent
We then have

\[
{{\partial {\bf P}} \over {\partial  x}} \propto {\bf r}_f
\]

\noindent
where

\[
{\bf  r}_f  =  \left(\rho,  c_f,  -\frac{c_f  B_x  B_y}{\Delta_f},  0,
\frac{\rho c^2_f B_y}{\Delta_f}\right)^t \mbox{~~with~~} \Delta_f=\rho
c_f^2 -B_x^2.
\]

\noindent
is  the right  eigenvector for  a  fast-mode wave  propagating in  the
positive  $x$  direction  for  the  system written  in  terms  of  the
primitive  variables,  ${\bf  P}$.   Note  that  we  assume  that  the
fast-mode wave is non-degenerate i.e.  neither $B_y$ nor $B_x$ vanish.

Like FH02, we use the  above expression to superpose a small amplitude
sinusoidal fast-mode wave propagating in the positive $x$-direction on
a  uniform  background  at  rest.   In  order to  make  the  flow  two
dimensional, we  also introduce a  shift with respect to  the $x$-axis
that depends on $y$. The initial state is given by

\begin{equation}
{\bf  P}(x,y,0) =  {\bf P}_0  + A_1\sin\left\{\frac{2\pi}{L_1}\left[x+
A_2\sin\left(\frac{2\pi y}{L_2}\right)\right]\right\}{\bf r}_f,
\label{eq:initial}
\end{equation}

\noindent
where $A_1$ and $L_1$ are the amplitude and wavelength of the fast-mode  
wave and $A_2$  and $L_2$  the amplitude  and wavelength  of the
phase  shift.  The homogeneous  background state  ${\bf P}_0$  is as
in FH02:

\[
\rho = 1, {\quad} p=T_{\rm e}\rho, {\quad} {\bf v}=0, {\quad} B_{x}=1,
{\quad} B_{y} = \alpha B_{x}.
\]
 
Here  $T_{\rm e} (=  a^2)$  is  the  equilibrium temperature  of  the
isothermal gas.   As in FH02,  we use $T_{\rm  e} = 10^{-3}$ (or  $a =
0.0316$). Since the  amplitude is small, the fast-mode  ${\bf r}_f$ is
calculated using the values of the unperturbed state and is normalised
so that the amplitude of $v_{x}$ is $A_1$.

Density   inhomogeneities   are    generated   for   all   values   of
$\alpha$. However, the highest contrasts are obtained when the initial
wave  propagates  at a  modest  angle to  the  field  (FH02).  In  our
calculations we therefore  limit ourselves to $\alpha =  0.25$. We set
$A_1  = 0.05$  since the  transverse magnetic  field changes  sign for
larger amplitudes .  The non-linear  steepening of the wave then leads
to  the  formation  of  an  intermediate shock,  which  is  unphysical
\citep{FK01}. We adopt $A_2 = 1$ for the amplitude of the phase shift.
Other values for the model parameters are being considered in Sect.
\ref{sect:parameter study}. 

Note that in  the initial state, ${\bf P}(x,y,0)$,  $B_x$ is constant,
while $B_y$ is a function of $y$. This means that the divergence
constraint  $\nabla .  {\bf  B} =  0$  is violated.  However, for  the
adopted parameters (see also below), we  find that $\nabla . {\bf B} <
10^{-5}$, which  is of the same  order of magnitude  as the divergence
errors that arise in  numerical MHD calculations due to discretisation
of the equations.  As we include the divergence  cleaning algorithm of
\citet{D02} to  stabilize the numerical scheme,  the initial violation
of the divergence constraint is damped out.

Our initial state has an advantage over other models 
which introduce the two-dimensionality in a more natural way, i.e. a
simple $y$-dependence of the primitive variables that allows an unambiguous 
interpretation of the initial stages of the simulations. A small density 
variation or a magnetic field $B_x$ that varies slightly with $y$
introduces a complex $y$-dependence for the primitive variables (through 
$c_f$ and $\Delta_f$). Calculations done with these alternative initial 
states, however, produce similar results to those of our model and hence
provide a posteriori justification for our choice.

The computational domain is $0 \leq x \leq L_1$ and $0 \leq y\leq L_2$
($L_1 = L_2  = L =$ 1000) with periodic boundary  conditions. We use a
uniform grid with $1\,000 \times 1\,000$ mesh points.

\section{Results}

\label{sect:results}

\begin{figure}
\includegraphics[width = 8.4 cm]{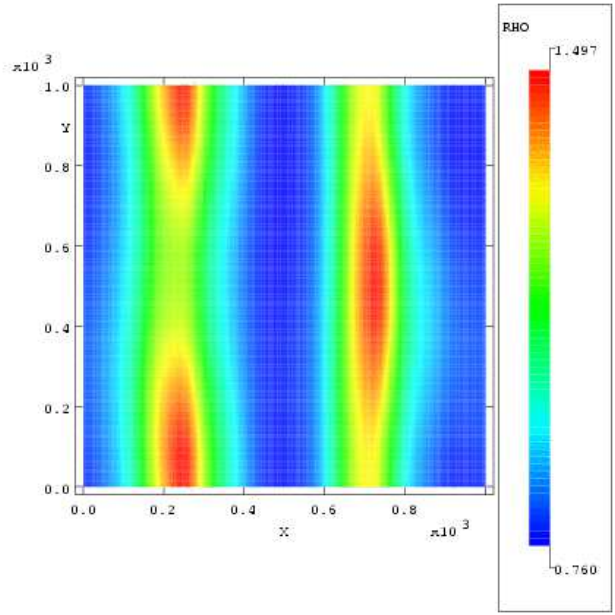}
\caption{The density  at t  = 2\,000. The  contour lines are  added to
show that  two regions can  be distinguished. The central  region lies
approximately within $250 < y < 750$.}
\label{fig:asymmetry}
\end{figure}

Initially the solution is almost one dimensional: the gradients in the
$y$-direction are of the order $L/2\pi A_2 (\approx 100)$ smaller than
the  gradients in  the  $x$-direction. Although  these variations  are
small, they  cannot be  neglected. From Eq.~(\ref{eq:initial})  we see
that the gradients in the  $y$-direction change sign when $y=L/4= 250$
and $y=3L/4  =750$ (for any given  value of $x$). This  means that the
domain  can  be  divided  into  two regions  with different  temporal
evolution: a  central part with $250  < y <  750$ and the rest  of the
domain.   In Fig.~\ref{fig:asymmetry}  we  can clearly  see these  two
regions. Note that  at this stage the variations  in the $y$ direction
are of the order of a few percent.

\begin{figure}
\includegraphics[width = 8.4 cm]{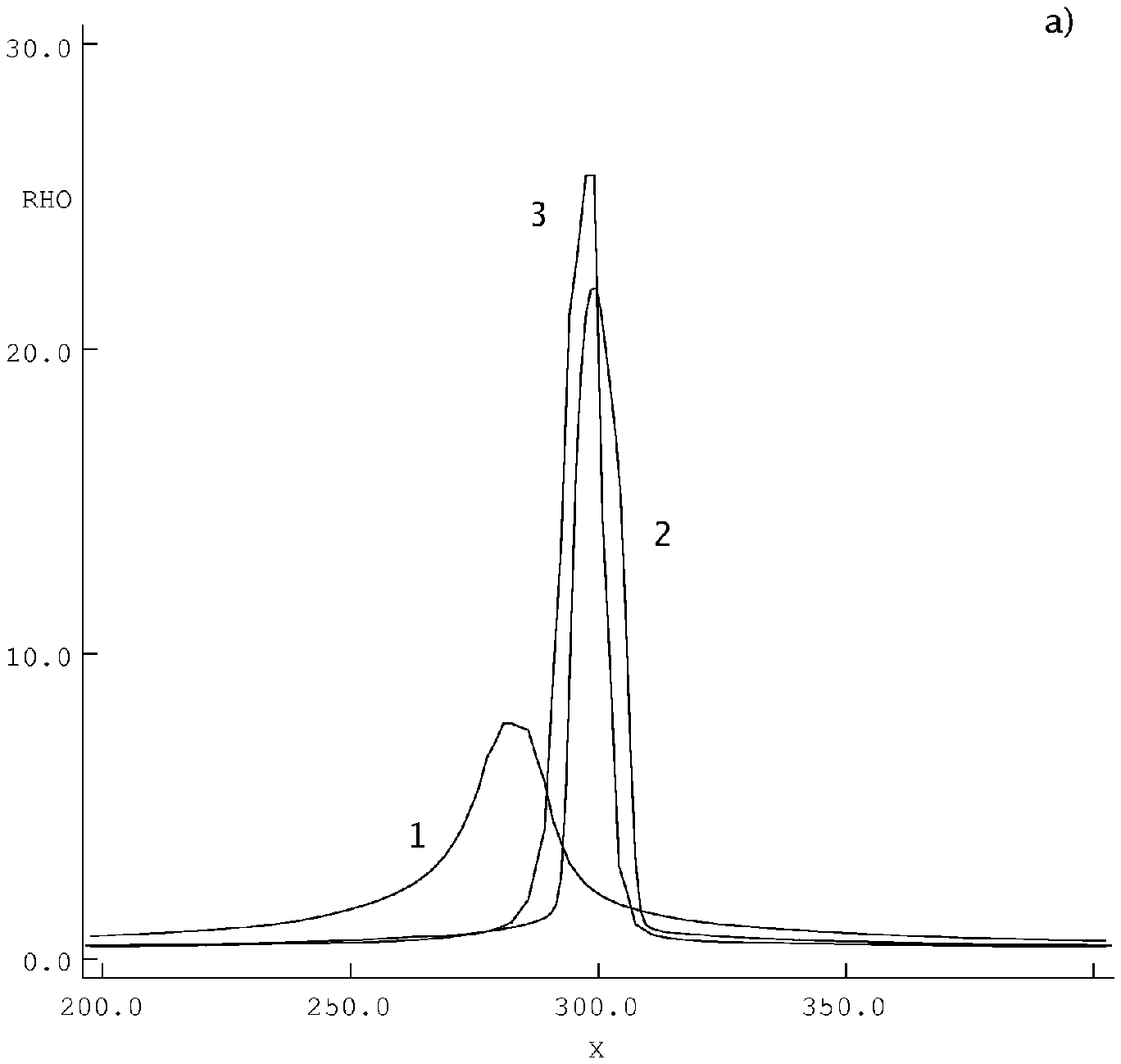}
\includegraphics[width = 8.4 cm]{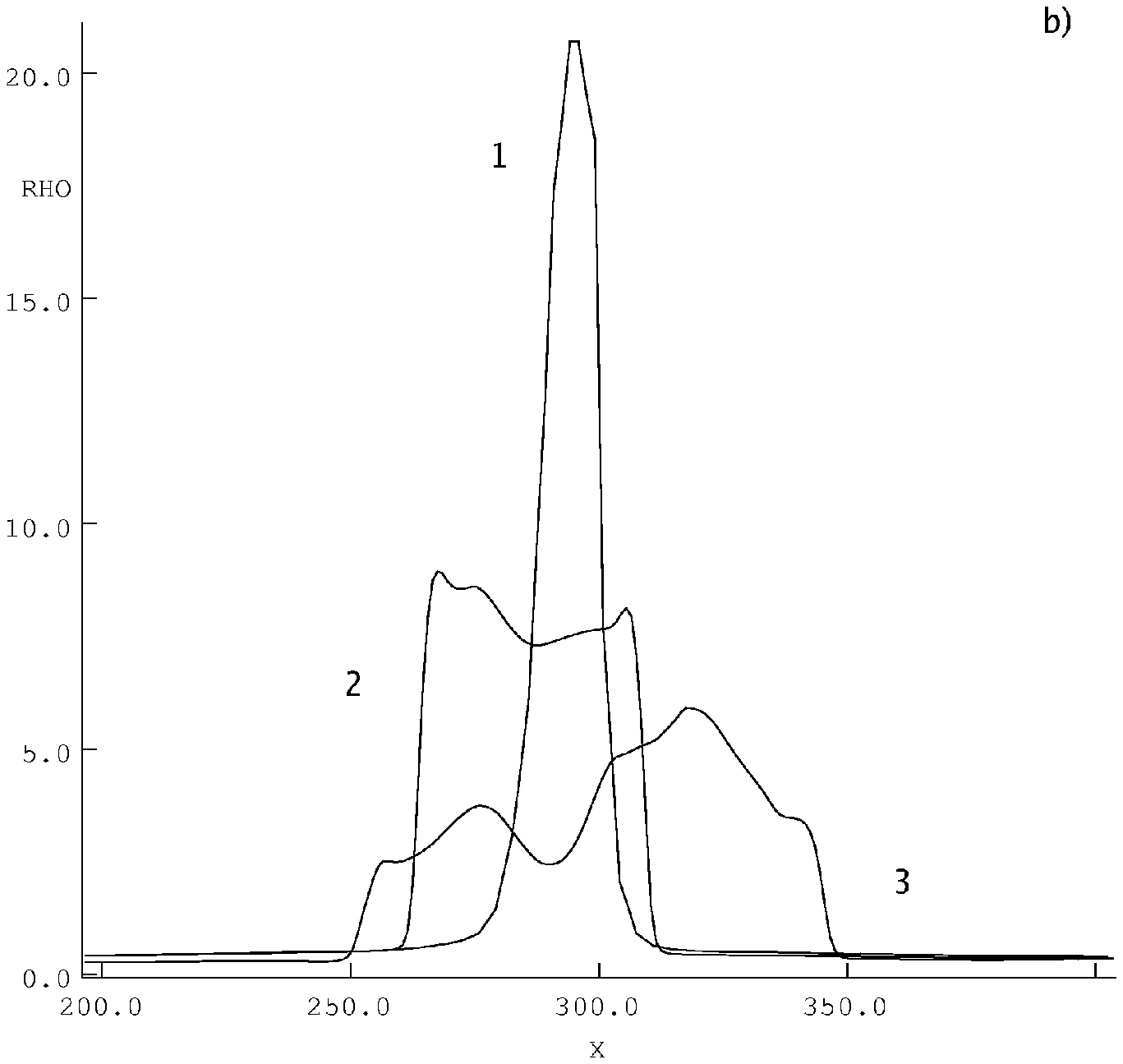}
\caption{(a) The growth of the  density perturbation at $y = 250$. The
times are $t = 3 \times 10^3$, 2 -  $t = 3.5 \times 10^3$ and 3 - $t =
3.6 \times 10^3$. (b) Decay of  the density perturbation at $y = 250$.
The times are 1 - $t = 3.7 \times 10^3$, 2 - $t = 4 \times 10^3$ and 3
- $t = 4.5 \times 10^3$.}
\label{fig:shock}
\end{figure}

\begin{figure}
\includegraphics[width = 8.4 cm]{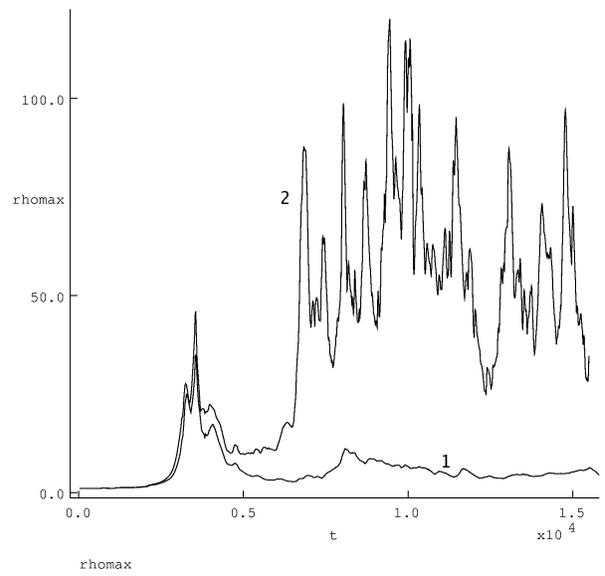}
\caption{The maximum density measured  in the computational domain for
both 1D (1; from FH02) and 2D (2) as a function of time.}
\label{fig:rho_history}
\end{figure}

\begin{figure}
\includegraphics[width = 8.4 cm]{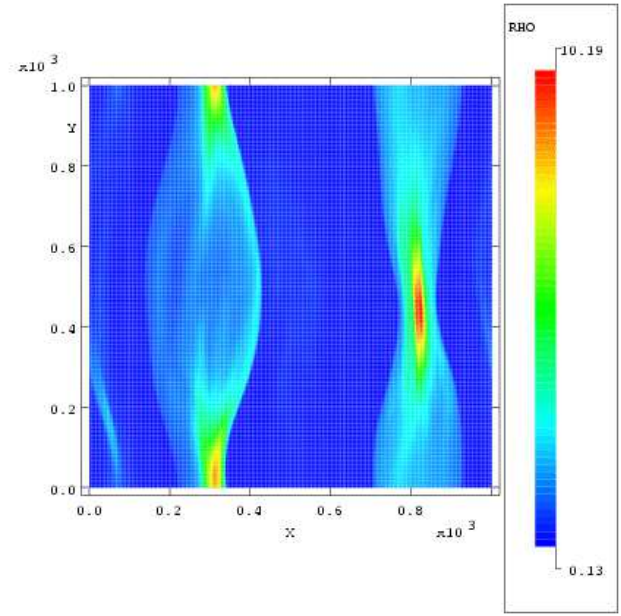}
\caption{The density  at $t =  5\,500$. The density  perturbations are
bounded by curved shocks. The density is highest in the central region
where the shocks are moving slowly.}
\label{fig:curved_shocks}
\end{figure}

After $3$  periods (i.e. $t \approx  3\, 000$) the  fast-mode steepens
into a shock, thereby  exciting slow-mode components. FH02 showed that
these slow modes produce inhomogeneities with large density contrasts.
Since there are as yet no significant variations in the $y$-direction,
these   high-density   perturbations   aligned  with   the   $y$-axis.
Figure~\ref{fig:shock} shows the growth of one such high-density 
region and its subsequent decay. Up to $t \approx 5\,000$, the maximum
densities in the computational domain do not differ significantly from
those in 1D (see  Fig.~\ref{fig:rho_history}). The density profiles of
the   perturbations   are  also   similar   to   the   1D  ones   (see
Fig.~\ref{fig:shock} and FH02).

We  can   also  see   from  Fig.~\ref{fig:shock}b  that   the  density
perturbation is bounded by shocks as it expands. In contrast to the 1D
case, the speed in the laboratory frame, $v_s$, of these shocks is not
constant along the shock front.  The upstream fluid speed in the frame
of  the shock,  $v_u$,  is roughly  constant  and equal  to $A_1  c_a$
(FH02), where $c_a =  |B_x|/\rho^{1/2}$ is the initial Alfv\'en speed,
whereas  the   upstream  speed  in  the  rest   frame,  $v_p$,  varies
considerably  along the  shock front.   Since $v_s  = v_p  -  v_u$, it
follows that $v_s$ must also vary along the shock front i.e. the shock
front   is   curved.    This   effect   can   be   seen   clearly   in
Fig.~\ref{fig:curved_shocks}.

\begin{figure}
\includegraphics[width = 8.4 cm]{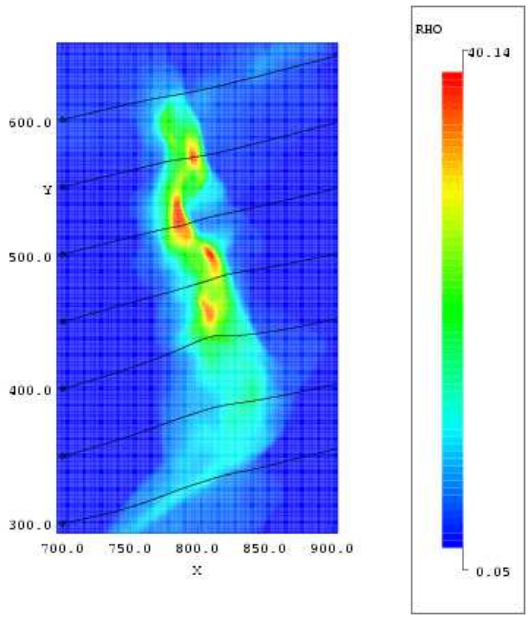}
\caption{Hierarchical  structure in  the  density at  t  = 12\,000:  a
number of dense cores reside in  a larger clump. The magnetic field is
represented by the solid lines.}
\label{fig:dense_cores}
\end{figure}

An important consequence of the curved shock front is the formation of
clumps, rather  than dense  sheets. The dense  regions are  bounded by
slow shocks that propagate predominantly in the $x$-direction. Regions
that  are bounded  by slowly  moving  shocks therefore  retain a  high
density  for longer  than those  bounded by  faster shocks.   At later
times this  produces variations in  the $y$-direction that  are larger
than those that occurred before  the initial decay.  However, there is
little motion  in the $y$-direction  because this is  prevented by the
magnetic  field. Once  a  clump  forms, its  long  axes is  orientated
perpendicular to the magnetic field lines, but its aspect ratio is not
particularly large (see Fig.~\ref{fig:dense_cores}).

\begin{figure}
\includegraphics[width = 7.2 cm]{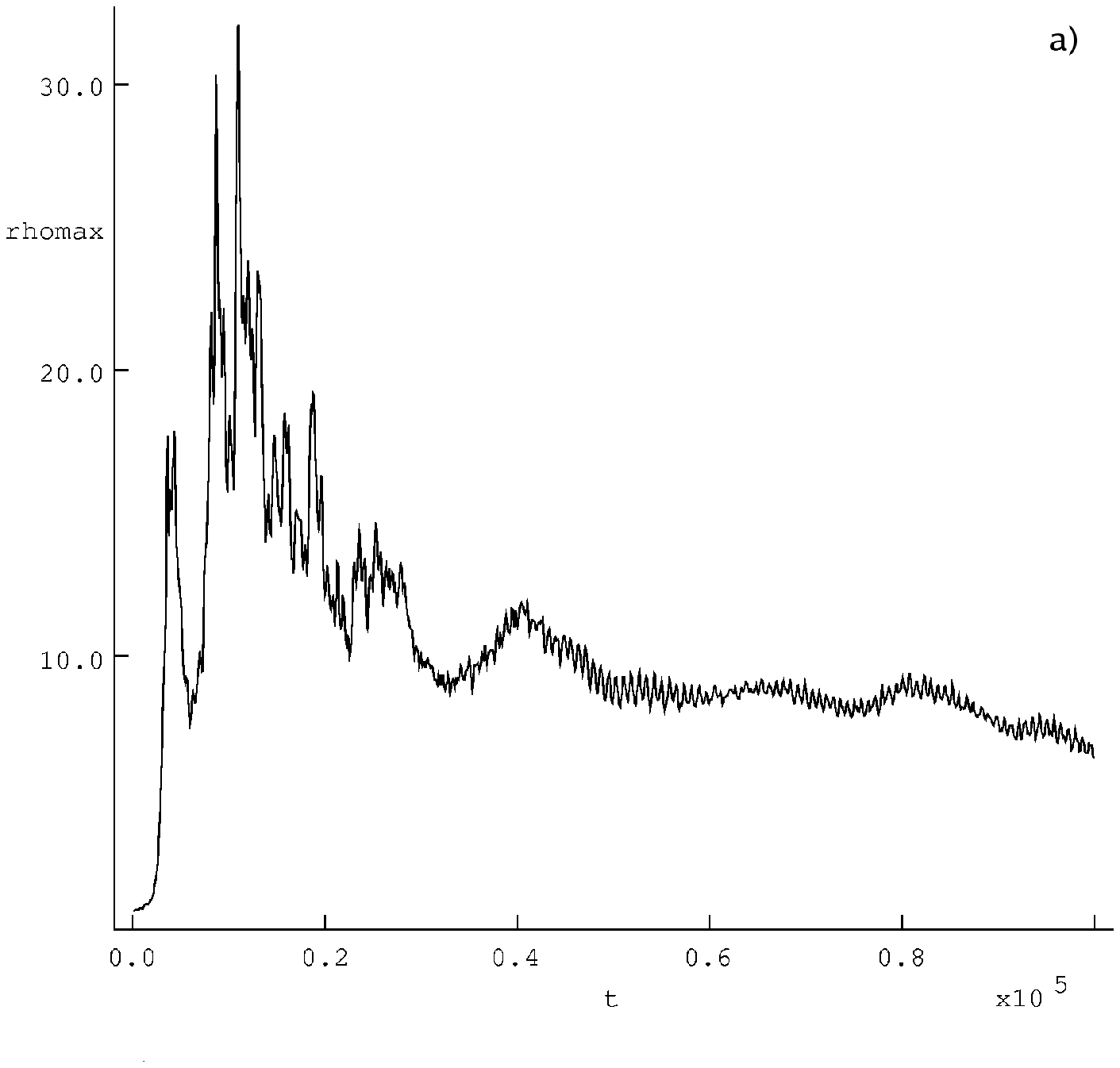}
\includegraphics[width = 7.2 cm]{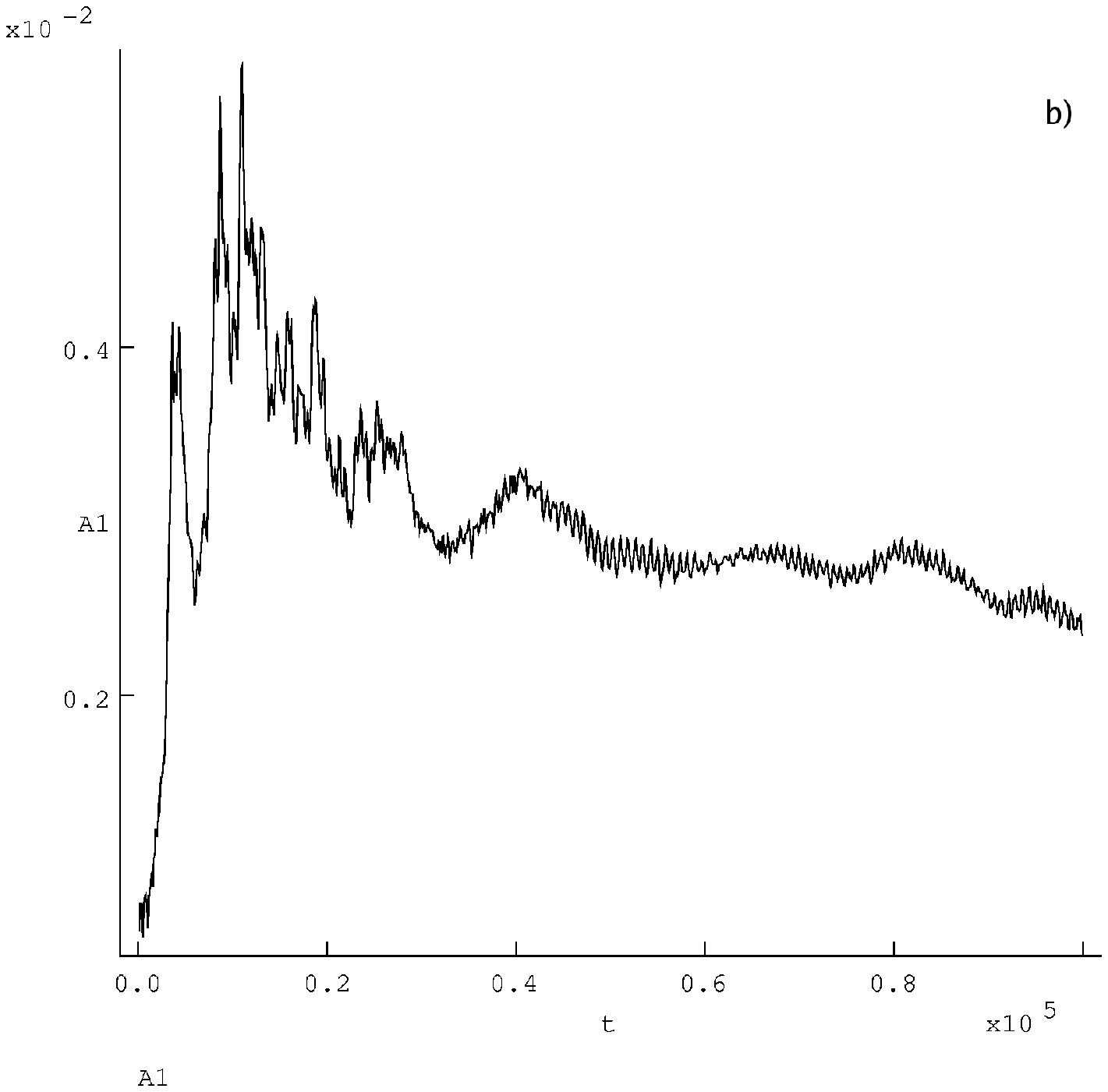}
\includegraphics[width = 7.2 cm]{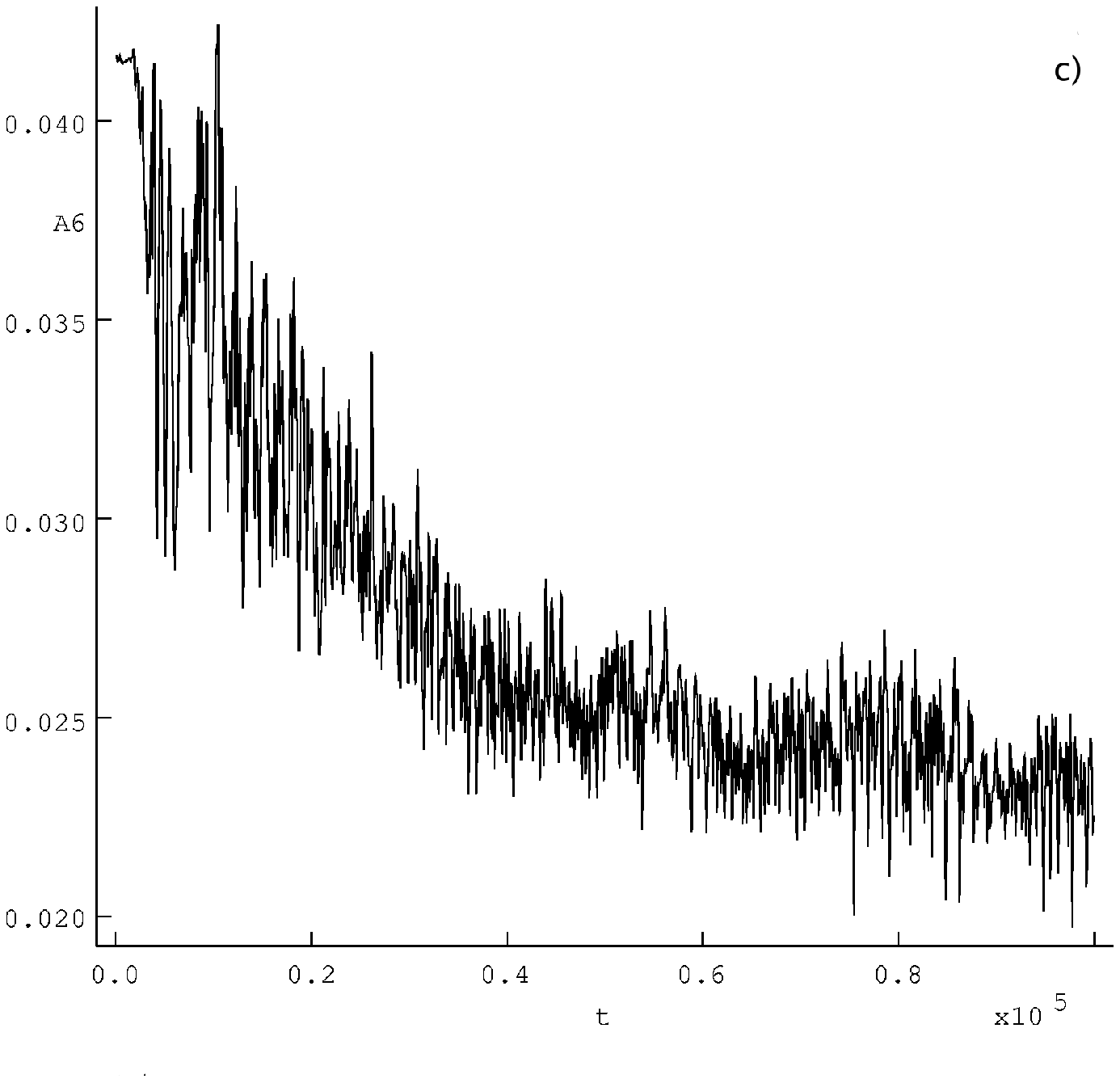}
\caption{(a) Similar to Fig.~\ref{fig:rho_history}, but now for
a  uniform  grid  with  $100  \times  100$ mesh  points  up  to  $t  =
100\,000$. (b)  The maximum value  for the amplitude-measure of the 
slow-mode wave in the computational domain (for definition see FH02). 
(c) Similar to (b), but for the fast-mode wave.} 
\label{fig:fast mode decay}
\end{figure}

Although  the average density  within clumps  is nearly  constant, the
maximum   density  can   have  large   fluctuations.   As   seen  from
Fig.~\ref{fig:rho_history}, the maximum densities  are a factor of two
higher  than obtained in  1D calculations.   The initial  fast-mode is
still present and  interacts with the clumps and  its bounding shocks,
thereby generating  slow-mode waves. 
Figure~\ref{fig:fast mode decay} shows a measure of the amplitude
of fast and slow-mode waves as defined in FH02. It shows clearly the  
correlation of the slow-mode waves with the density inhomogeneities. 
The slow modes are thus responsible for
the   fragmentation  of   the  clump   into   high-density  contrasts.
Figure~\ref{fig:dense_cores} shows  a close-up  of a clump  with dense
substructures.  Such fragmentation is, of course, not possible in 1D.

The  dense substructures  in the  clump  are thus  transient and  have
rather short lifetimes. They are generated because slow-mode waves are
continuously excited by the fast-mode wave interacting with the clump.
This explains why  the time-scales of the density  fluctuations are of
the order of the time for  a fast-mode wave to cross the computational
domain, i.e. $t =  1\,000$. As this wave dissipates, the amplitude of 
the slow modes decreases, resulting in lower  density  contrasts  (see
Fig.~\ref{fig:fast  mode  decay}).    Furthermore,  the  clump  itself
disperses. The 2D  results show that a clump persists  for a period of
time that exceeds the lifetimes  of the density perturbations in 1D by
a factor of 10.

The ratio, $\beta$, of the gas pressure to magnetic pressure increases
in our simulations from its  initial value of $\beta \approx 0.002$ to
$\beta \approx 0.2$. While the gas pressure can become a factor of 100
larger than initially  since it scales like the  density, the magnetic
field does not  vary much in strength. When  $\beta$ approaches unity,
the process  of producing high-density contrasts by  slow modes ceases
to  be effective. There  is therefore  an upper  limit to  the maximum
density that can be produced.

\section{Parameter study}

\label{sect:parameter study}

In the  previous section,  we focused on  one specific  numerical case
with   a   set   of   specific   parameters.  As   the   results   are
parameter-dependent,  we   briefly  discuss  the   influence  of  each
parameter.

\paragraph*{$\balpha$}{\bf :}
The fast-mode wave is  most efficient in exciting slow-mode components
for modest  values of  the angle, $\alpha$,  between its  direction of
propagation  and the  magnetic  field (FH02).   When  $\alpha$ is  too
large, the  shock is not effective in  producing slow-mode components,
whereas  there  is  little  steepening  when $\alpha$  is  too  small.
Moreover, the fast-mode wave is also responsible for the excitation of
slow modes  within  the  clumps.   Since  it is  the  slow modes  that
generate  the density  inhomogeneities,  this means  that models  with
$\alpha \neq 0.25$ produce lower density contrasts.

\paragraph*{${\bf A_1}$ and ${\bf a}$}{\bf :}
Both parameters affect the  maximum density produced by the non-linear
steepening  of  the fast-mode  wave.   FH02  showed  that the  maximum
density is roughly  proportional to $A^2_1/a^2$.

While  the  time-scale  for  the  formation  of  the  substructure  is
independent of  $a$, it depends on  $A_1$ in two ways  (see FH02). The
wave amplitude determines  the time for the fast-mode  wave to steepen
into a shock and also the time-scale for the subsequent decay. A lower
value of $A_1$ means that dense cores arise later in the simulations.

\begin{figure}
\includegraphics[width = 8.4 cm]{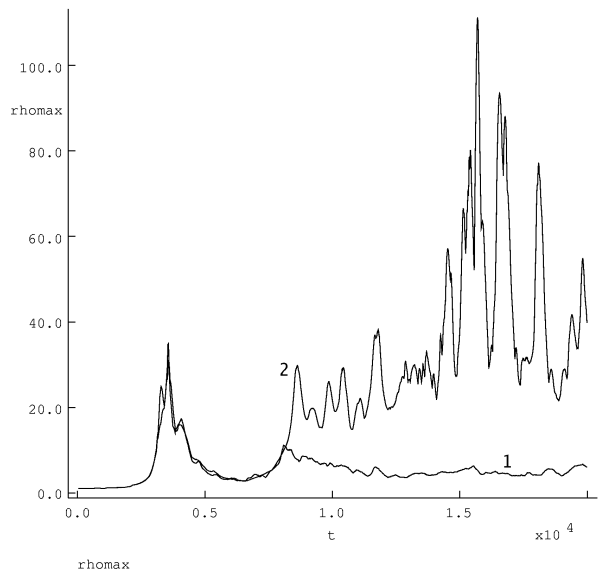}
\caption{Similar  to Fig.~\ref{fig:rho_history},  but now  for  $A_2 =
0.1$.}
\label{fig:rho_history2}
\end{figure}

\paragraph*{${\bf A_2}$}{\bf :}
The  phase-shift  of  the   fast-mode  wave  is  responsible  for  the
variations along the $y$-axis.  The amplitude of this shift determines
the   time    at   which   the    simulations   cease   to    be   1D.
Figure~\ref{fig:rho_history2} shows the  time evolution of the maximum
density for  an amplitude of $A_2  = 0.1$.  As  the initial variations
are a  factor of 10  smaller than  for $A_2 =  1$, it takes  longer to
develop any significant  deviation from 1D.  This also  means that the
formation of  substructure is deferred. Note that  the maximum density
that    can   be    attained    is   independent    of   $A_2$    (see
Fig.~\ref{fig:rho_history} and Fig.~\ref{fig:rho_history2}).

\paragraph*{${\bf L_1}$ and ${\bf L_2}$}{\bf :}
The wavelengths  of the fast-mode  wave and the phase  shift determine
the  dimensions of  the clump.   Its extent  in the  $x$-direction is
roughly comparable with the  size of density perturbations produced by
excitation  of  slow modes.   This  means  that  it  is  $\approx  L_1
\rho_0/\rho_{\rm  max}$, where  $\rho_0$  is the  initial density  and
$\rho_{\rm max}$ the  maximum density.  The size in  the $y$-direction
is  set  by  the wavelength  of  the  phase  shift.  As  mentioned  in
Sect.~\ref{sect:results}, a sinusoidal  phase shift divides the domain
into two separate regions so that the $y$-extent is roughly $L_2/2$.

\section{Discussions and conclusions}

\label{sect:discussions and conclusions}

We  have extended  the FH02  model which  describes the  generation of
density inhomogeneities by MHD waves  in a low-$\beta$ environment, to
two dimensions.   We have followed  the evolution of a  fast-mode wave
propagating  in   the  positive   $x$-direction  with  a   small,  but
non-negligible,  variation  in the  $y$-direction  produced  by a  
$y$-dependent  shift along  the  $x$-axis.  Like FH02,  we  find that  the
formation of density inhomogeneities is associated with the excitation
of slow modes by the fast-mode wave.

An important  result is  that the process  of generating  structure by
slow  modes  works  on  different  length-scales.   While  the  larger
structures arise  due to the  non-linear steepening of the  fast mode,
there are smaller structures generated by the interaction between
the  fast-mode wave  and the  larger structures.   A  limiting factor,
however, is that $\beta$  increases within each level of substructure.
When $\beta$  approaches unity, the  mechanism ceases to  be effective
and  produces no  further  substructures.  $\beta$  is probably  about
unity in dense cores (e.g. \citealt{M90}).

Although we made no attempt  at reproducing any observational data, we
can investigate  the relevance of  our simulations to  observations of
clumps and dense  cores in GMCs. The structure  of molecular clouds is
often described  as filamentary or clumpy  (e.g.  \citealt{BW97}). For
example,  observations of  the $\rho$  Oph A  core show  an arc-shaped
ridge  with 4  embedded cores  \citep{AWB93}.  Similar  structures are
obtained in our  simulations. However, the shape of  the clump depends
on the properties of the  initial fast-mode wave, i.e. its wavelength,
and  on   the  length-scale   of  perpendicular  variations   (in  our
calculations determined by the sinusoidal  phase-shift of the fast-mode
 wave).  Other  shapes  of  clumps  can then  be  reproduced by 
MHD waves with  different properties.

\citet{BS86} found that the clumps in the Rosette Molecular Cloud, an 
archetypal GMC, are an order of magnitude denser than the mean cloud 
density. Dense cores within the clumps typically have a density that  
exceeds the mean density by two orders of magnitude, but peak 
densities in pre-stellar cloud cores can range to values that exceed 
the  mean density by 4 orders of magnitude (e.g. \citealt{L03}). In 
our simulations, we find that the average density contrasts in the 
larger structures are in excess of 10 times the average background  
density of 1. For the substructures, i.e. individual dense cores, this 
increases to 100. We do not find the observed peak values, but this can
be readily explained by the fact that we have restricted ourselves to 
modest amplitudes of the initial fast-mode wave. Also we have not
included self-gravity in our model.  The inclusion of this effect will
be   investigated  in   a  subsequent   paper.   However,   we  expect
self-gravity only to act as an  amplifier and not to have an influence
on the fragmentation of the molecular cloud.

Recently,  \citet{G05} investigated the  chemistry of  transient dense
cores within molecular  clouds. Using the results of  FH02 to describe
the  growth and  decay of  the dense  cores, they  find  a significant
enhancement   in  chemical   compositions   of  the   cores  as   they
disperse.  This  leads  eventually   to  chemical  enrichment  of  the
molecular  cloud.  In our  simulations,  however, high-density  clumps
survive  much longer  than they  do in  FH02.  Our  results  thus have
serious implications for the chemical evolution and richness of clumps
and dense cores.

\section*{Acknowledgments}
We thank the anonymous referee for a report that helped to 
improve the manuscript.
SVL gratefully acknowledges PPARC for the financial support.

\label{lastpage}

\end{document}